\documentstyle[aps,psfig,floats]{revtex}
\begin{document}
\draft
\title{Correlations in the cotunneling regime of a quantum dot}  
\author{Reinhard Baltin $^1$ and Yuval Gefen $^2$}
\address{$^1$ Max Planck Institut f\"ur Kernphysik, Postfach 103980,
  69029
  Heidelberg, Germany\\
  $^2$ Department of Condensed Matter Physics, The Weizmann Institute
  of Science,
  Rehovot 76100, Israel} 
\date{\today}
\maketitle
\begin{abstract}
Off--resonance conductance through weakly coupled quantum dots 
("valley conductance") is governed by cotunneling processes in which a
large number of dot states participate. Virtually the same states
participate in the transport at consecutive valleys, which leads to
significant valley-valley conductance correlations. These correlations are
calculated within the constant interaction model. Comparison with
experiment shows that these correlations are less robust in reality. Among
the possible reasons for this is the breakdown of the constant interaction
model, accompanied by "scrambling" of the dot as the particle number is
varied.
\end{abstract} 
\pacs{PACS numbers: 73.20.Dx, 73.23.Hk, 73.40.Gk} 
\section{Introduction}
One of the interesting aspects of the physics of quantum dots is 
the mechanism of cotunneling \cite{averinnazarov} which governs 
transport through the quantum dot away from resonance ("conductance valley"). 
Such a mechanism, which usually gives a small (as compared with 
resonance values) yet significant contributions to the conductance,  
consists of the use of a large number of virtual dot states, which, 
due to high electrostatic energy, are classically forbidden. As one 
varies an external parameter (e.g. applied magnetic field), these are 
virtually the same (albeit possibly modified) states that contribute 
to the transmission, giving rise
to significant conductance correlations \cite{aleinerglaz96}.
In a recent work \cite{sumrule} we have pointed out that such correlations
show up in the transmission 
{\em phase} 
as well. Here we study the
conductance at different valleys (corresponding to a different number of
electrons on the dot, a parameter which is controlled by an applied gate
voltage) and calculate the ensuing conductance correlations. Our
formalism, outlined in the present and throughout the next section,
follows that of Aleiner and Glazman \cite{aleinerglaz96}.  We evaluate the
significant valley--valley conductance correlations within the constant
interaction model, as function of the strength of the interaction,
location in the valley and temperature (Section III and Appendix A). A
simplified toy model which, in our opinion, captures much of the pertinent
physics, is presented and studied in Appendix B. Comparison with the
experiment (Section IV) reveals that in reality these correlations are
less robust. Possible reasons for this are discussed in Section V, where
we stress the likelihood of the breakdown of the constant interaction
model, accompanied by "scrambling" of the dot states as the particle
number is varied.

We consider transport through a quantum dot weakly coupled to reservoirs 
by high tunneling barriers \cite{vanhouten,curacao}. The chemical 
potential $\mu$ of the quantum dot
can be tuned by an applied gate voltage $V_g$. Because of the small
size of those structures charging effects are important; the weak
coupling to the external reservoirs (leads) implies that, in general, 
there is a definite number of electrons on the dot. 
The conductance of the
quantum dot as function of $V_g$ shows nearly equally spaced
peaks. 
These resonances are due to sequential tunneling of electrons
through the dot. At these values of $V_g$ the ground state energies of
the system with the dot containing $N$ and $N+1$ electrons are
degenerate and
the occupation number can fluctuate. In the ``valleys''
between the resonances the conductance is strongly suppressed due to
the charging energy: it costs energy to add or remove electrons to or
from the dot, and the number of electrons on the dot is fixed to an
integer value. 

There is however a residual conductance in the valley between the
peaks which is caused by a different transfer process. An electron or
hole tunnels virtually from one reservoir to the other through
energetically forbidden states. Since the particle tunnels coherently
through the whole structure this transfer mechanism is known as
cotunneling \cite{averinnazarov}. In the following we will focus on
this regime away from the resonances.

We model the quantum dot coupled to the reservoirs by the usual 
Hamiltonian
\begin{eqnarray}
H &=& H^L+H^R+H^{QD}+H^T \ , \\
H^{L(R)} &=& \sum_{k} \varepsilon_k a_k ^{L(R) \dagger} a_k ^{L(R)}\ ,
\\
H^{T} &=& \sum_{k,j} V_{j,k} ^L c_j ^{\dagger} a_k ^L + h.c. +
L \leftrightarrow R \ , \\
H^{QD} &=& \sum_{j} (\epsilon_j-\mu) c_j ^{\dagger} c_j +\frac{U}{2}
\hat{N}(\hat{N}-1). 
\label{ham}
\end{eqnarray}
$H^{L,R}$ describe the reservoirs to the left and right of the
quantum dot, $H^T$ represents the tunneling of electrons in and out 
of the quantum dot,
and $H^{QD}$ describes the states of the isolated quantum dot
including the
electron-electron interaction. We use the simplifying ``constant
interaction'' model which asserts that the total energy due to the
electron-electron interaction 
solely depends on the total number of electrons on the dot ($\hat{N}$
is the number operator). The
charging energy is given by $U=e^2/C$ invoking the electrostatic
energy of a classical
capacitance $C$. The coupling 
strength of level $j$ of the quantum dot
to the leads is characterized by the tunneling rates 
$\Gamma_j=2\pi/\hbar \sum_k |V_{j,k}|^2 \delta (E-\epsilon_k)$.

For the calculation of transport properties we need the retarded Green
function of the quantum dot coupled to the leads,
\begin{eqnarray}
G^{\rm ret}_{i,j}(t)&=& \langle [c_i (t),c^{\dagger}_j(0) ] _+ \rangle \ .
\label{gretres}
\end{eqnarray}
In the regime of weak coupling $\Gamma\ll kT, \Delta$ 
(temperature $T$, mean level spacing $\Delta$) we approximate
$G^{\rm ret}$ by 
\begin{eqnarray}
G^{\rm ret}_{j,k} &=& \sum_{N=0}^{\infty} P_N \delta_{jk} \left [
\frac{\langle n_j \rangle_N}{E-(\epsilon_j -\mu+ U\cdot (N-1))
  +i\Gamma_j} \right. 
+\left. \frac{1-\langle n_j \rangle_N}{E-(\epsilon_j-\mu
+ U \cdot N) +i\Gamma_j} \right ].
\label{green}
\end{eqnarray}
with $\langle ... \rangle_N =\mbox{tr}_N \exp -\beta H^{QD}
.../\mbox{tr}_N \exp -\beta H^{QD}$ the thermal average with $N$
electrons. Here $\Delta$ is the single particle level spacing. The 
probability to find $N$ electrons on the dot is given
by $P_N=\langle \hat{N} \rangle_N/\sum_M \langle \hat{N}
\rangle_M$. Away from the resonances eqn. (\ref{green}) describes the 
elastic cotunnling,
i.e. the virtual tunneling via a single state $j$ of the quantum
dot. This is the dominant process in the regime $kT<\sqrt{U\Delta}$
\cite{averinnazarov} which we are focussing at.    
In addition, (\ref{green}) also describes the dynamics in the vicinity
of the resonances on the same level as an master equation approach of
ref. \cite{beenakker}.

For the Green function $G^{\rm ret}$ we note that in the cotunneling regime
away from the resonances there is an integer number $\bar{N}$ of electrons
on the dot, and $P_N=\delta_{N,\bar{N}}$ in (\ref{green}). 
Hereafter we attach an 
index $N$ to a valley, corresponding to the number of electrons on the
dot over that range of $V_g$. The resonance separating the valleys
$N-1$ and $N$ will be denoted by $(N-1,N)$. It is convenient to 
shift the reference point of the chemical potential $\mu$
to the resonance $(N-1,N)$ ($\mu \rightarrow \mu-\epsilon_N -
U(N-1)$), and make use of the parametrization of $\mu$ by $x$:
for the valley $N$ the parameter $x\rightarrow0$
corresponds to the right of resonance $(N-1,N)$, i.e. the point where the
energies of the system with $N-1$ electrons and $N$ electrons on the
dot are
practically degenerate; $x\rightarrow1$ corresponds to the left of resonance
$(N,N+1)$. A complete description of $\mu$ is given by the two 
variables $(N,x)$.

Eqn. (\ref{green}) contains canonical occupation numbers. 
In the following we
will use the grand-canonical occupation numbers, i.e. the Fermi
functions, instead. We account for the difference by using an
effective inverse temperature $\beta_{GC}\approx \beta(1+\beta \Delta/4)$ 
\cite{kamenev} which is justified for $\beta \Delta \ll 1$. 
We then obtain with $\omega=\epsilon_j$
\begin{eqnarray}
G^{\rm ret}_{\omega}(N,x) &=& \frac{f_{\beta_{GC}}(\omega-\epsilon_N)}
{\epsilon_N-\omega
  +xU}- \frac{f_{\beta_{GC}}(\epsilon_N-\omega)}{\omega-\epsilon_N
  +(1-x)U}.
\label{gretzero}
\end{eqnarray}
Since we are in the cotunneling regime (large denominators), we
neglect the tunneling rates
$i\Gamma$ in the denominators in eqn. (\ref{green}).
The first (second) term describes the occupied (empty)
states with energies smaller (greater) than $\epsilon_N$.
The minimum energy necessary for hole (particle) transfer is  
$E_h=x U$ ($E_p=(1-x)U$) with $E_h+E_p=U$. 
We should keep in mind that
eqn. (\ref{gretzero}) is valid for $x$ sufficiently away from the
resonances at $x=0,1$. 
Since the number of electrons $N$ starts to 
fluctuate on an energy scale $kT$ around the resonances 
the range of validity of eqn. (\ref{gretzero}) is 
$kT/U < x < 1-kT/U$. 

\section{Formalism}
\label{formal}

We will briefly review the formalism of Aleiner and Glazman
\cite{aleinerglaz96} for the calculation of 
transport properties of a disordered quantum dot in the
cotunneling regime.
Hereafter ``disorder'' should be understood as caused either by
impurity
scatterers or by shape irregularities in the confining potential.
In both cases the statistical properties of the
single-particle levels are described by random matrix theory (RMT)
(for a review see \cite{therev}): over energy intervals smaller than 
the Thouless energy $E_{Th}$,
the tunneling matrix elements and the level spacing are strongly 
fluctuating.  

We start with the expression for the transmission
amplitude $t(E)$ containing the retarded Green function of the dot
$G^{\rm ret}$ and the tunneling matrix elements $V^{L,R}_j$, 
\begin{eqnarray}
t(E) &=& \sum_{j} V^L_j(E) V^{R *}_j(E) G^{\rm ret}_{j}(E) \ ,
\label{telast}
\end{eqnarray}
where $V^L_j=V^L_{j,k(E)}\sqrt{2\pi \rho^L (E)}$ with $\rho^L$ the
density of states in the left reservoir.
Since in the elastic cotunneling regime the electrons or holes are 
transferred through one single dot state, only the diagonal elements
in (\ref{gretres}) have to be accounted for.
The tunneling matrix elements $V_{j,k(E)}$ in the Hamiltonian
(\ref{ham}) contain the overlap of the lead wave function with the dot
wave function at the barrier \cite{bardeen}. We assume that the 
tunneling matrix
elements factorize into a part describing the tunnel barrier with its
penetration factor, and a dot part. The dot part involves the wave
function of state $j$ at the barrier, $\psi_j(R)$. Thus, we have for
the left barrier $V_{j,k(E)}^L=V^L \psi_j(R^L)$. The separation is
justified when the properties of the barrier only weakly change on
the typical energy scale, $U$, of the system. Eqn. (\ref{telast}) then
becomes 
\begin{eqnarray}
t(E)&=& 2 \pi \sqrt{\rho^L \rho^R} V^L V^{R *} 
\sum_{j} \psi_j(R^L) \psi_j ^* (R^R) G^{\rm ret}_j(E) \nonumber\\
&=& 2\pi \sqrt{\rho^L \rho^R} V^L V^{R *} 
\int d\omega \sum_{j} \psi_j(R^L) \psi_j ^* (R^R) 
\delta (\omega-\epsilon_j) G^{\rm ret}_{\omega}(E) \ .
\label{tdelt}
\end{eqnarray}
with $\rho^{L,R}$ the density of states in the leads.
For the last step in (\ref{tdelt}) it is essential that $G^{\rm ret}_j$ only
depends on $j$ via $\epsilon_j$, cf. (\ref{gretzero}). Aleiner and
Glazman identified the sum over $j$ as the local density of states of
single-particle levels and expressed it in terms of Green functions of
the non-interacting dot
\begin{eqnarray}
G^A_{\omega}(R^L,R^R)-G^R_{\omega}(R^L,R^R)&=& \sum_{j} \left[
\frac{\psi_j(R^L) \psi^{*}_j(R^R)}{\omega -\epsilon_j- i\delta}- 
\frac{\psi_j(R^L) \psi^{*}_j(R^R)}{\omega -\epsilon_j+ i\delta} \right]
\nonumber \\
&=&
2\pi i \sum_{j} \psi_j(R^L) \psi^{*}_j(R^R) \delta(\omega-\epsilon_j).
\label{localdos}
\end{eqnarray}
Impurities on the quantum dot only 
affect the single-particle states.
With (\ref{localdos}) the transmission amplitude is expressed as a
convolution of two types of Green functions,
\begin{eqnarray} 
t(E) &=& 2\pi \sqrt{\rho^L \rho^R} V^L V^{R *} 
\int_{}^{} \frac{d\omega}{2\pi i} (G^A_w(R^L,R^R)-
G^R_{\omega}(R^L,R^R)) G^{\rm ret}_{\omega}(E) \ .
\label{separate}
\end{eqnarray}
One term in the integral is the cotunneling Green function 
$G^{\rm ret}$ of a quantum dot
with interaction (but no
explicit dependence on disorder dependent quantities); the other term
includes the advanced and retarded Green functions $G^A$ and $G^R$ of the
system which account for disorder, but do not include interaction. We
would like to stress that this separation into interaction and disorder
dependent terms respectively is only possible
due to the constant interaction model.

The transmission amplitude $t(E)$ is linked with
the linear conductance ${\cal G}$ through the quantum dot by the 
Landauer formula \cite{landauer70,meirwingreen}. 
Off the resonances 
${\cal G} =e^2/h |t(E_F)|^2$ even for finite temperature.
Averaging over disorder only affects the single-particle Green
functions $G^{R,A}$ defined for non-interacting electrons.
For weak disorder, $k_F l\gg1$, a diagrammatic expansion can be
established \cite{abrikosov}.
The averages $\langle G^R G^R\rangle$ and $\langle G^A G^A \rangle$
reduce to the product of the average of the single particle Green
functions; as long as the dot's size $L=|R^L-R^R|$ is larger than the
mean free path $l$ such terms may be neglected.
The remaining average $\langle G^R G^A \rangle$ can be
expressed in terms of Diffusons and Cooperons,
\begin{eqnarray} 
\langle G^R_{\omega_1, B_1}(R^L,R^R) G^A_{\omega_2, B_2}(R^R,R^L) 
\rangle &=& 2\pi \rho^D {\cal D}_{\omega_1-\omega_2}^{B_1,B_2},\\
\langle G^R_{\omega_1, B_1}(R^L,R^R) G^A_{\omega_2, B_2}(R^L,R^R) 
\rangle &=& 2\pi \rho^D {\cal C}_{\omega_1-\omega_2}^{B_1,B_2}
\end{eqnarray}
where $\rho^D$ is the density of states on the dot. Diffusons
(Cooperons) depend
only on the difference (sum) of magnetic fields $B_1, B_2$, and are
the solution of diffusion equations. In momentum space
($R^L\leftrightarrow k$, $R^R\leftrightarrow \bar{k}$) these
solutions read
\begin{eqnarray}
\label{solndiff}
{\cal D}_{\omega}^{B,B} (q)&=& \frac{S^{-1}}{-i\omega +Dq^2} \ , \\
{\cal C}_{\omega}^{B,B} (Q)&=& \frac{S^{-1}}{-i\omega +DQ^2 +\Omega^+}
\ ,
\label{solncoop}
\end{eqnarray}
where $q=k-\bar{k}$ and $Q=k+\bar{k}$. The area of the quantum dot is
denoted by $S$ and $\Omega^+ =B^2 E_{Th} S^2/\Phi_0^2$ with $\Phi_0=hc/e$
the flux quantum. The Cooperon contribution is
maximal for $Q=0$, i.e. for the coherent backscattering 
process $k=-\bar{k}$. In
contrast to the Diffuson, the Cooperon is suppressed for a
magnetic field greater than a characteristic field $B_c\sim \Phi_0/S$. The
lowest non-vanishing eigenvalues of the diffusion operators 
are of the order of
the Thouless energy $E_{Th}$. For energies $E<E_{Th}$ only the
zero-frequency mode must be retained which corresponds to spatially
homogeneous solutions in (\ref{solndiff}) and (\ref{solncoop})
($q,Q\rightarrow 0$).      

We focus on the regime $\Delta \ll E_h,E_p <
E_{Th}$. The first inequality assures that we are allowed to use the
diffusion approximation. Since the relevant energy
scale (charging energy) is smaller than the Thouless energy $E_{Th}$ 
which is the case for a dot with not too strong disorder or boundary
deformations, we only have to consider the spatially
homogeneous zero mode $q\rightarrow0$. 
The sum of Diffusons in momentum space then becomes
\begin{eqnarray}
& &{\cal D}_{\omega}(q)+{\cal D}_{-\omega}(-q) =
\frac{S^{-1}}{-i\omega+Dq^2}+ \frac{S^{-1}}{i\omega+Dq^2}  \nonumber \\
& & = -i S^{-1} \left (\frac{1}{\omega-i Dq^2}- \frac{1}{\omega+i Dq^2}
\right ) 
\stackrel{q\rightarrow 0}{\longrightarrow} 2 \pi S \delta(\omega).
\label{diffdelta}
\end{eqnarray}
Aleiner and Glazman have obtained the full distribution
function of ${\cal G}$  
\cite{aleinerglaz96} and have found for the variance
\begin{eqnarray}
\mbox{var} {\cal G} &=& \langle {\cal G}^2 \rangle -\langle {\cal G}
 \rangle^2 = \left \{ \begin{array}{ll} 2 \langle {\cal G} \rangle^2 &
 \mbox{for } B=0 \\ \langle {\cal G} \rangle^2 & \mbox{for } B\gg B_c
\end{array} \right. \ .
\label{varian}
\end{eqnarray}

\section{Cotunneling conductance correlations}

In \cite{aleinerglaz96} the conductance-conductance correlation for
different magnetic fields but the same chemical potential was
studied. Here, we will calculate 
$\langle {\cal G} (N_1,x_1) {\cal G} (N_2,x_2) \rangle$
between different values of the chemical potential
$(N_1,x_1)$ and $(N_2,x_2)$
but fixed magnetic field for the two cases $B=0$ and $B\gg B_c$.
We are especially interested in the correlation function for
$N_2=N_1+\Delta N$ and $x_1=x_2$, i.e. for different valleys but at
the same position within the valleys.    
In order to make contact to the experiment \cite{cronenwett} 
we consider the following normalized autocorrelation function
\begin{eqnarray}
C(N_1,x_1,N_2,x_2) &=& \frac{\langle {\cal G}(N_1,x_1) 
  {\cal G}(N_2,x_2) \rangle- \langle {\cal G}(N_1,x_1)\rangle \langle
  {\cal G}(N_2,x_2) \rangle}{\sqrt{\mbox{var}{\cal
  G}(N_1,x_1)}\sqrt{\mbox{var} {\cal G}(N_2,x_2)}}
\label{autocorr}
\end{eqnarray}
for $U< E_{Th}$. We expect to find strong correlations since 
from one valley to the next only one level changes its character from
occupied to empty (or vice versa) \cite{sumrule}. For instance,
the first empty state in the $N^{th}$ valley is the last occupied
state in the $(N+1)^{st}$ valley. 

\begin{figure}[h!]
\begin{tabular}{lll}
\begin{minipage}{5.7cm}
\centerline{
\psfig{file=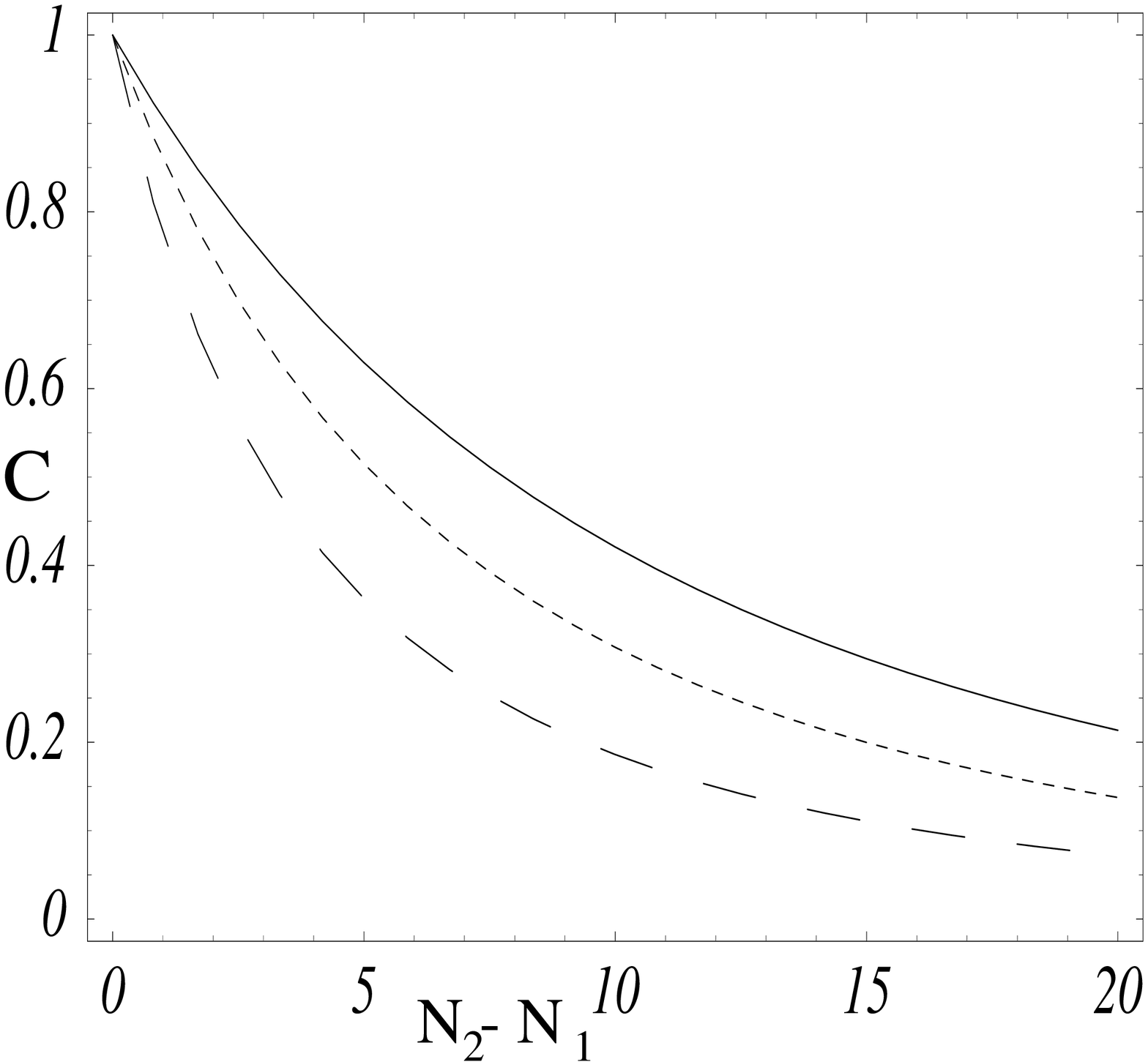,height=6cm,width=5.7cm,angle=0}}
\end{minipage}
&
\begin{minipage}{5.7cm}
\centerline{
\psfig{file=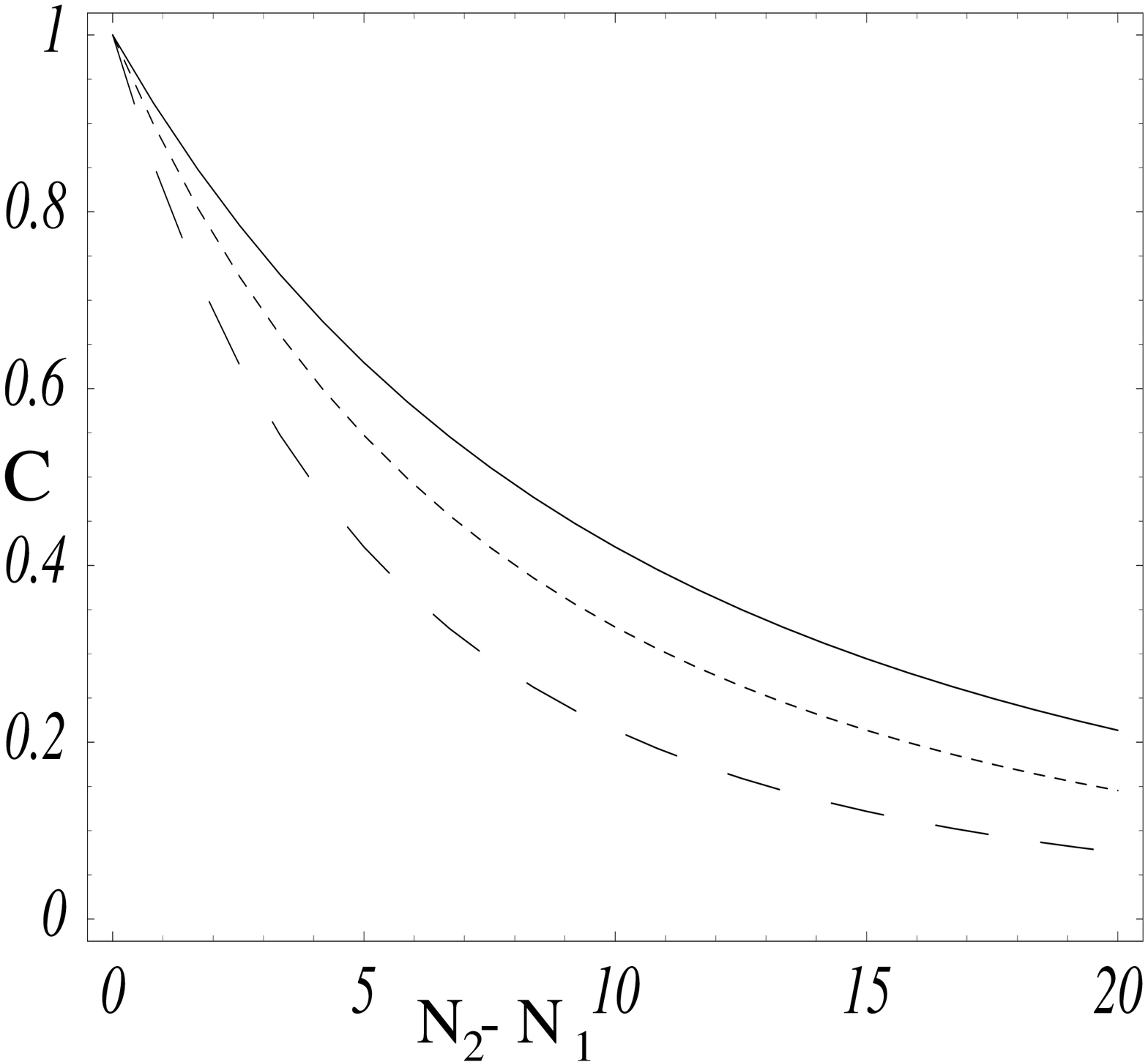,height=6cm,width=5.7cm,angle=0}}
\end{minipage}
&
\begin{minipage}{5.7cm}
\centerline{
\psfig{file=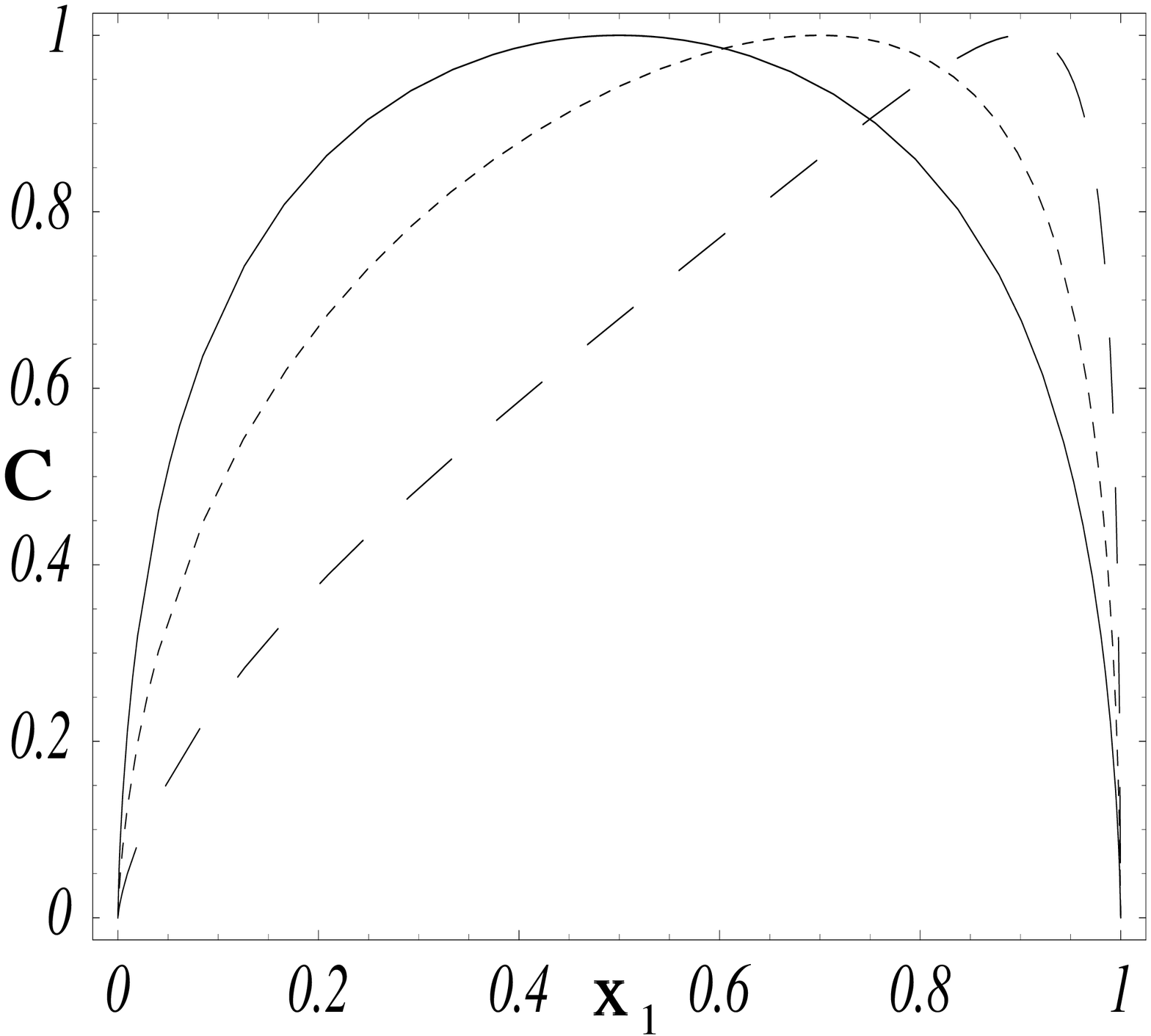,height=6cm,width=5.7cm,angle=0}}
\end{minipage}
\end{tabular}
\caption{Conductance autocorrelation function $C$ for $x_1=x_2=x$
  vs. valley number 
  $N_2-N_1$. Left: For $U/\Delta=40$ and $x=0.5$ (solid), $0.2$
  (dotted) and $0.1$ (dashed). Center: For $x=0.5$ and $U/\Delta=40$
  (solid), $30$ (dotted) and $20$ (dashed). Right: $C$ for $N_1=N_2$
  vs. relative position in the valley $x_1$ and $x_2$ fixed at $0.5$
  (solid), $0.7$ (dotted) and $0.9$ (dashed). Note that
  (\protect{\ref{correlx1x2}}) is not valid in the vicinity of order
  $kT/U$ around $x_1=0$ and $x_1=1$.}
\label{condcorr1}
\end{figure}

We start as shown in the previous section by
separating interaction from disorder
\begin{eqnarray}
\lefteqn{\langle{\cal G}(N_1,x_1) {\cal G}(N_2,x_2) \rangle =
{\cal K}^2 \int d\omega_{1} 
d\omega_2 d\omega_3 d\omega_4 \times \nonumber }\\ 
& & G^{\rm ret}_{\omega_1}(N_1,x_1) G^{\rm ret}_{\omega_2}(N_1,x_1)^* 
G^{\rm ret}_{\omega_3}(N_2,x_2) G^{\rm ret}_{\omega_4}(N_2,x_2)^*
\times \nonumber \\
& &
\left\langle 
\left (G^A_{\omega_1}(R^L,R^R)-G^R_{\omega_1}(R^L,R^R) \right)
\left (G^R_{\omega_2}(R^R,R^L)-G^A_{\omega_2}(R^R,R^L) \right)
\right. \times \nonumber \\
& & \left.
\left (G^A_{\omega_3}(R^L,R^R)-G^R_{\omega_3}(R^L,R^R) \right)
\left (G^R_{\omega_4}(R^R,R^L)-G^A_{\omega_4}(R^R,R^L) \right) 
\right \rangle \ .
\label{condcorrel}
\end{eqnarray}
with ${\cal K}=e^2/h |V^L|^2 |V^R|^2 \rho^L \rho^R$.
Since the energy scale ($U$) involved in this problem is much larger than
the mean level spacing ($\Delta$) the average of the product of four
Green functions can approximately be decoupled into a product of pairwise
averaged Green functions \cite{aleinerglaz96}. In a diagrammatic
expansion this approximation corresponds to keeping a certain class
of diagrams which also turn out to be relevant for the universal
conductance fluctuations \cite{altshuler,leestone85}.  
As in the previous section we will neglect terms like 
$\langle G^R G^R\rangle$ and $\langle G^A G^A\rangle$. All possible
pairings of the remaining averages in terms of Diffusons and Cooperons
are given by 
\begin{eqnarray}
\lefteqn{\langle (G^A-G^R)(G^A-G^R)(G^A-G^R)(G^A-G^R) \rangle = \nonumber}\\ 
& & \left ({\cal D}_{\omega_2-\omega_1}(R^R,R^L)+
{\cal D}_{\omega_1-\omega_2}(R^L,R^R) \right) 
\left ({\cal D}_{\omega_4-\omega_3}(R^R,R^L)+
{\cal D}_{\omega_3-\omega_4}(R^L,R^R) \right) + \nonumber\\
& & \left ({\cal D}_{\omega_4-\omega_1}(R^R,R^L)+
{\cal D}_{\omega_1-\omega_4}(R^L,R^R) \right) 
\left ({\cal D}_{\omega_2-\omega_3}(R^R,R^L)+
{\cal D}_{\omega_3-\omega_2}(R^L,R^R) \right) +\nonumber\\
& & \left ({\cal C}_{\omega_3-\omega_1}(R^L,R^R)+
{\cal C}_{\omega_1-\omega_3}(R^L,R^R) \right) 
\left ({\cal C}_{\omega_2-\omega_4}(R^R,R^L)+
{\cal C}_{\omega_4-\omega_2}(R^R,R^L) \right) \ .
\label{spliteqn}
\end{eqnarray}
Due to the Cooperons, the conductance-conductance
correlation function (\ref{condcorrel}) depends on the magnetic field.
Using (\ref{spliteqn}), eqn. (\ref{condcorrel}) 
can be expressed as
\begin{eqnarray}
\lefteqn{\langle{\cal G}(N_1,x_1) {\cal G}(N_2,x_2) \rangle =
\langle {\cal G}(N_1,x_1) \rangle \langle {\cal G}(N_2,x_2) \rangle +
\nonumber }\\
& & + \left | 2\pi \rho^D {\cal K} \int d\omega_{1} d\omega_2  
G^{\rm ret}_{\omega_1}(N_1,x_1) G^{\rm ret}_{\omega_2}(N_2,x_2)^* 
\left( {\cal D}_{\omega_2-\omega_1} (R^R,R^L)+ {\cal
    D}_{\omega_1-\omega_2} (R^L,R^R) \right) \right |^2 + \mbox{\hspace{1em}} 
\nonumber \\
& & + \left | 2\pi \rho^D {\cal K} \int d\omega_{1} d\omega_2  
G^{\rm ret}_{\omega_1}(N_1,x_1) G^{\rm ret}_{\omega_2}(N_2,x_2) 
\left( {\cal C}_{\omega_2-\omega_1} (R^L,R^R)+ {\cal
    C}_{\omega_1-\omega_2} (R^L,R^R) \right) \right |^2 \ .
  \mbox{\hspace{1em}} 
\end{eqnarray}
For a strong magnetic field the last term involving the Cooperons
vanishes.
For $U<E_{Th}$ we only have to consider the spatially constant zero mode
of the diffusion operator, and the sum of the two Cooperons 
also tends to $\delta(\omega_1-\omega_2)$ for $B=0$, as in
(\ref{diffdelta}). Thus the conductance correlation function becomes
\begin{eqnarray}
\lefteqn{\langle{\cal G}(N_1,x_1) {\cal G}(N_2,x_2) \rangle =
\langle {\cal G}(N_1,x_1) \rangle \langle {\cal G}(N_2,x_2) \rangle +
 \alpha (4\pi^2 \rho^D {\cal K}S^{-1})^2  
| A(N_1,x_1,N_2,x_2) |^2 \ , 
\nonumber }\\
& & \mbox{where \hspace{1em}} A(N_1,x_1,N_2,x_2) =   
\int d\omega  
G^{\rm ret}_{\omega}(N_1,x_1) G^{\rm ret}_{\omega}(N_2,x_2)^*. 
\mbox{\hspace{4em}} 
\label{defa}
\end{eqnarray} 
The parameter $\alpha$ depends on whether time reversal symmetry is
broken or not: $\alpha=2$ at $B=0$, while $\alpha=1$ for $B\gg B_c$.
Thus the conductance correlation function (\ref{autocorr}) becomes
\begin{eqnarray}
C(N_1,x_1,N_2,x_2) &=&
\frac{|A(N_1,x_1,N_2,x_2)|^2}{|A(N_1,x_1,N_1,x_1)|
  |A(N_2,x_2,N_2,x_2)|}.
\end{eqnarray}
We note that $C$ is independent of $\alpha$, i.e. possesses a
universal form for both conserved and broken time reversal symmetry. 
The integral for $A$ at finite temperature 
can be evaluated and the result is given in
Appendix \ref{statea}. 
At low temperature $\Delta \beta \gg 1$ the integration yields  
with $d=\epsilon_{N_2}-\epsilon_{N_1}$ positive 
\begin{eqnarray}
\lefteqn{A(N_1,x_1,N_2,x_2) =
  \frac{1}{d+(x_2-x_1)U} \log\frac{d+x_2 U}{x_1 U} 
+ \frac{1}{d+(x_2-x_1)U} \log \frac{d+(1-x_1)U}{(1-x_2)U}
- \nonumber } \\
& & \mbox{\hspace{1em}}-\frac{1}{d+U+(x_2-x_1)U}\left( \log
  \frac{d+x_2 U}{x_2 U} +  
  \log \frac{d+(1-x_1)}{(1-x_1)U}\right) \ , \mbox{\hspace{6em}} 
\label{ttstar}
\end{eqnarray}
The autocorrelation
function (\ref{autocorr}) in this limit finally becomes
\begin{eqnarray}
C(N_1,x_1,N_2,x_2) &=&
x_1 x_2 (1-x_1)(1-x_2) \left[ \frac{1}{d/U+x_2-x_1} \log
\frac{(d/U+x_2)(d/U +1-x_1)}{x_1(1-x_2)} \right. -
\nonumber \\
& & \left. -\frac{1}{d/U+ 1 + x_2-x_1} \log 
\frac{(d/U +1 -x_1)(d/U+x_2)}{(1-x_1)x_2} \right]^2 \ .
\label{endresult}
\end{eqnarray}
The energy difference $d$ between the single-particle levels $N_2$ and
$N_1$ can be expressed in terms of the mean level
spacing $\Delta$, $d/U=(N_2-N_1)\Delta/U=(N_2-N_1)\delta$ where
$\delta=\Delta/U$.     
As a special case (and for the sake of simplifying the algebraic
expressions), we set $x=x_1=x_2$ and obtain for the conductance 
autocorrelation function $C$ for different valleys
\begin{eqnarray}
& & C(N_1,x,N_2,x) = x^2 (1-x)^2 \times \mbox{\hspace{8cm}}\nonumber \\
& &\lefteqn{
  \times\left[\frac{1}{(1+(N_2-N_1)\delta)(N_2-N_1)\delta}\log
  (1+\frac{(N_2-N_1)\delta}{x})(1+\frac{(N_2-N_1)\delta}{1-x})
  \right]^2} \ .
\label{correln1n2}
\end{eqnarray}
Fig. \ref{condcorr1}(left) shows $C$ as function of $N_2-N_1$ for
$U/\Delta=40$ and different values of $x$. As expected the 
correlations decay slowly with valley number.
Away from $x=1/2$, $C$ falls off more rapidly because
the nearest occupied/empty level contributes more
strongly to the conductance. Fluctuations in this level from one
valley to the next reduce the correlations.
Fig. \ref{condcorr1} (center) demonstrates that the scale for the 
decay is set by
$U/\Delta$. The slope of $C$ at $N_2-N_1=0$ at $x=0.5$ is given by
$-4\Delta/U$. 
  
Next we turn to the correlations between $x_1$ and $x_2$ in the same
valley $N_1=N_2$. The autocorrelation function (\ref{endresult})
then becomes
\begin{eqnarray}
C(N,x_1,N,x_2) &=& \frac{x_1 x_2 (1-x_1)(1-x_2)}{(x_2-x_1)^2} 
\left( \log\frac{x_2}{x_1}\frac{1-x_1}{1-x_2} \right)^2 \ .
\label{correlx1x2}
\end{eqnarray}
We note that $C$ is independent of the ratio $U/\Delta$. Fig.
\ref{condcorr1} (right) shows $C$ vs. $x_1$ and $x_2$ fixed. 
At $x_1=x_2$ the autocorrelation function becomes maximal, 
and it falls rapidly off for
$x_1\rightarrow 0,1$. We should bear in mind, however, that
(\ref{gretzero}) breaks down near the resonances, since it neglects
fluctuations in the particle number of the dot.  

\section{Comparison with an experiment}

Measurements of the inter-valley conductance of
semiconductor quantum dots were first reported in
\cite{cronenwett}. The experimental set-up was similar to other
experiments reported in ref. \cite{marcus96,marcus98}.
The dots were all in the ballistic regime and an
irregular shape of the confining potential renders the motion of the
electrons chaotic, so that RMT describes the statistics of the
mesoscopic fluctuations in the sample. Additional gate electrodes
could distort the shape of the dot, and allow an ensemble average to
be obtained from a single sample. The experiment focused on the
conductance-conductance correlations in the elastic cotunneling regime
between different magnetic fields but at the same chemical potential, 
as calculated in \cite{aleinerglaz96}. Since
$kT<\sqrt{U\Delta}$ the observed cotunneling was indeed elastic. In
ref. \cite{cronenwett} the
dot was more strongly coupled to the leads (barrier conductance ${\cal
  G}^{L,R}\sim e^2/h$) than in \cite{marcus96,marcus98}, since otherwise the
cotunneling current would be very small and difficult to measure. Thus
the tunneling rate $\Gamma \sim 0.7\Delta$ and the averaged
conductance at the peaks is eight times larger than the cotunneling
conductance in the valleys ($\langle{\cal G}\rangle \sim 0.05e^2/h$).

\begin{figure}[h!]
\centerline{\psfig{file=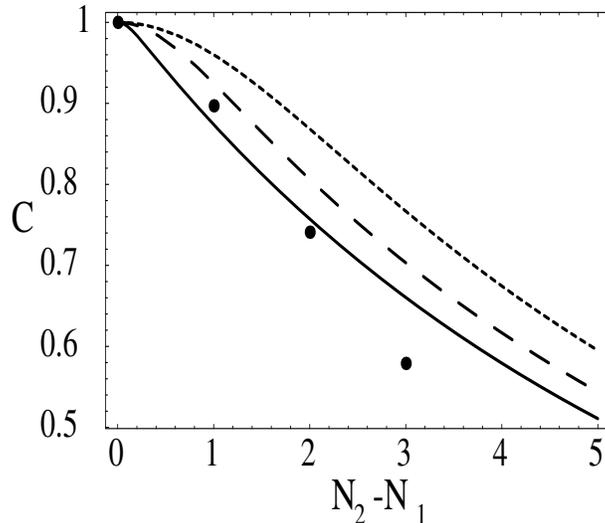,height=7cm,width=8cm,angle=0}}
\caption{Comparison between experiment (filled circles) and theory
  for the autocorrelation function $C$ at $x=1/2$ for
  several neighboring valleys ($U/\Delta=25.9$). Solid line:
  obtained from (\protect{\ref{correln1n2}}) at $\beta \Delta\gg 1$;
  dashed line: obtained from the expressions for $A$ from appendix
  \protect{\ref{statea}} at $\beta_{GC}\Delta=4$ which corresponds 
  to $\beta\Delta=2.47$;
 dotted: $\beta_{GC}\Delta=1.89$ corresponding to $\beta \Delta=1.4$.}
\label{compare}
\end{figure}

Fig. \ref{compare} shows a comparison of $C$ vs. $N_2-N_1$ between
theory and experiment (ref. \cite{cronenwett}). The quantum dot on the
sample under study had
the parameters 
$kT\sim 9\mu eV$, $\Delta=15.8\mu eV$ and $U=410\mu eV$.  
Experimental data for the autocorrelation function $C$
(\ref{autocorr}) in the middle of the
valleys $x_1=x_2=1/2$ are only available for three
neighboring valleys and for $U/\Delta=25.9$. The lines are our 
results for three different
temperatures. We observe that in the experiment the correlations are
more strongly suppressed than predicted in theory. 

\section{Discussion}

There are two main messages arising from the present analysis. On the
theoretical side, our analysis underlines the significant correlations of
the transmission (i.e. conductance) through a quantum dot as an 
external parameter
is varied. In our case the quantum dot is weakly coupled to the leads 
(external reservoirs) and the external parameter which is varied is the gate
voltage, affecting the number of electrons which reside on the dot. All
this results in valley-valley correlations of the conductance in the
cotunneling (far--from--resonance) regime. 
Previously, correlations have been found in the original analysis of 
Aleiner and Glazman \cite{aleinerglaz96} (intra--valley, as function 
of the magnetic field), in ref. \cite{kaminski} (valley--valley) 
considering the differential capacitance of the dot instead of the
cotunneling conductance, and in a recent analysis on transmission phases 
\cite{sumrule}. 
The mere existence of correlations as found in the
present study is, therefore, not totally surprising, although the details
(dependence on the location within the valley, on the interaction
strength--charging energy, temperature etc.) are certainly different.

The second message contains, in our opinion, a much more intriguing
element. The correlations found in the 
experiment seem to be less robust than our theoretical 
expressions suggest.
There might be several possible reasons for this.
Some are obvious while others are more subtle and may 
give rise to some intriguing physics.
\begin{itemize}
\item[(i)] 
The precise value of the electron-gas temperature in the experiment. In
\cite{cronenwett} the temperature was estimated to be $kT\sim9\mu eV$
which corresponds to $\beta\Delta=1.4$.
Modest deviations from this value are not going to produce agreement with
theory.
\item[(ii)] 
Approximating the canonical distribution function by a Fermi function
with an effective shifted temperature  (cf. the comment that precedes
eqn. (\ref{gretzero}), see also ref. \cite{kamenev}) is asymptotically 
justified for $kT\gg\Delta$,
which, for  $\beta \Delta =1.4$ is not quite the case. However, given the
other curves in Fig. \ref{compare}, a more accurate treatment of this 
point is not going to cure the problem.
\item[(iii)] 
In our derivation of the correlation function $C$ we have considered the
case $U< E_{Th}$. From the technical point of view, this allowed us to retain
the "zero-mode" of the Diffuson propagator only. In the experiment, the
Thouless energy was estimated to be $E_{Th}=180\mu eV$, which is smaller than
the charging energy of $U=410\mu eV$.  
Qualitatively one can expect 
that the inclusion of
non-zero modes would imply deviation from the constant interaction model.
In that case the addition (removal) of an electron is likely to change the
effective (interaction induced) potential landscape felt by the other
electrons, facilitating more efficient "scrambling" of the dot (see
below), and suppressing the correlations found here. However a
quantitative statement in this regard calls for a detailed analysis, not
included in the present work.
\item[(iv)] 
The Hamiltonian of an interacting electron system (such as a quantum dot)
includes terms other than the constant interaction $U$. Varying the gate
voltage is then bound to change the
nature of the many-body wave-function of the electron gas in the dot. For
example, within the approximate Hartree-Fock picture, the effective single
particle wave functions will be modified as electrons are kept added to
the dot, resulting in the breakdown of the Koopmans picture
\cite{koopman}.
It was shown that the Koopmans picture, which 
asserts that the effective
single-particle Hartree-Fock states remain unchanged as electrons are added
to or removed from the system, breaks down when it concerns with finite
(and disordered) systems, with sufficiently strong electron-electron  
interaction. For the systems studied that breakdown happened at $r_s \le
1.5$ where $r_s=\sqrt[3]{3/(4 \pi n_0 {a_0}^3)}$ with $n_0$ the
electron density and $a_0$ the Bohr radius. 
A remarkable experimental evidence has been provided by the
experiment of \cite{monkeysmile}, where the 
magnetofingerprints of 
various excited states of
a dot at different particle number have been compared. It turned out
that by adding 
electrons to the dot of concern 
(the total number of electrons was 200) the magneto fingerprints have
been significantly
modified, indicating scrambling of the electronic states. A more
systematic study of this scrambling has been reported in
ref. \cite{marcus98}. We speculate that the suppression of the
conductance correlations in
the present context may be another tool to evaluate the scrambling. One
systematic measurement which is called for is to repeat the experiment
for
different values of $r_s$. Evidently, to facilitate a more detailed 
comparison with our calculated expression one would need data concerning
various values of $U/\Delta$, varying temperature and a larger number of
valleys.

\section{Acknowledgment}
We acknowledge useful discussions with S.~M. Cronenwett, C.~M. Marcus, and
H.~A. Weidenm\"uller. After completing this work we learned that 
A. Kaminski, I.~L. Aleiner, and L.~I. Glazman calculated the
valley--to--valley correlation function of the differential
capacitance in the low temperature regime using a similar
approach \cite{kaminski}. We thank these authors for turning our 
attention to their work. This work was supported by the DIP
foundation, by the Center of Excellence of the Israel Science
Foundation founded by the Israel Academy of Sciences and Humanities, 
and by the Minerva Foundation.
\end{itemize} 

\begin{appendix}
\section{$A$ for finite temperature}
\label{statea}
For the conductance correlation function we need the integral  
\begin{eqnarray}
A(N_1,x,N_2,x)&=& \int d\omega \left[
\frac{f(\omega-\epsilon_N)}{\epsilon_N-\omega+x U+ i\Gamma}-
\frac{f(\epsilon_N-\omega)}{\omega-\epsilon_N +(1-x)U-i\Gamma}\right]
\times \nonumber\\
&\times & \left[
\frac{f(\omega-\epsilon_N-d)}{\epsilon_N+d-\omega+x U- i\Gamma}-
\frac{f(\epsilon_N+d-\omega)}{\omega-\epsilon_N-d +(1-x)U+i\Gamma}\right]
\end{eqnarray}
where $d=\epsilon_{N_2}-\epsilon_{N_1}$.
Since we are primarily interested in the correlation between different
valleys we set $x_1=x_2=x$. By means of contour integration we obtain
\begin{eqnarray}
A&=& \frac{1}{d(1-\exp(-\beta d))}
[\psi(\frac{1}{2}-\hat{\beta}(xU+d))-\psi
(\frac{1}{2}-\hat{\beta}xU)+
\psi(\frac{1}{2}-\hat{\beta}((1-x)U+d))-\psi
(\frac{1}{2}-\hat{\beta}(1-x)U)]+\nonumber\\
&+& 
\frac{1}{d(1-\exp(\beta d))}
[\psi(\frac{1}{2}-\hat{\beta}xU)-\psi
(\frac{1}{2}-\hat{\beta}(xU-d))+
\psi(\frac{1}{2}-\hat{\beta}(1-x)U)-\psi
(\frac{1}{2}-\hat{\beta}((1-x)U-d))-\nonumber\\
&-& \frac{1}{(d+U)(1-\exp(-\beta d))}
[\psi(\frac{1}{2}-\hat{\beta}(d+xU))-
\psi(\frac{1}{2}+\hat{\beta}(1-x)U) +
\psi(\frac{1}{2}-\hat{\beta}(d+(1-x)U))-
\psi(\frac{1}{2}+\hat{\beta}xU)]- \nonumber\\
&-& \frac{1}{(U-d)(1-\exp(\beta d))}
[\psi(\frac{1}{2}-\hat{\beta}(xU-d))-
\psi(\frac{1}{2}+\hat{\beta}(1-x)U) +
\psi(\frac{1}{2}-\hat{\beta}((1-x)U-d))-
\psi(\frac{1}{2}+\hat{\beta}xU)]
\end{eqnarray}
where $\psi$ is the Digamma function and $\hat{\beta}=\beta/(2\pi i)$. 
For the normalization in (\ref{autocorr}) we need $A$ for
$d\rightarrow 0$. We obtain
\begin{eqnarray}
A(N,x,N,x)&=& i\left( \frac{\beta}{2\pi}-\frac{1}{\pi U} \right)
\left(\psi^{(1)}(\frac{1}{2}-\hat{\beta}U(1-x))+\psi^{(1)}(\frac{1}{2}-
\hat{\beta} U x) \right) -\nonumber \\
&-& \frac{\beta}{4\pi^2} \left( 
\psi^{(2)}(\frac{1}{2}-\hat{\beta}U(1-x))+\psi^{(2)}(\frac{1}{2}-
\hat{\beta} U x) \right) 
\end{eqnarray}
with $\psi^{(1,2)}$ the first and second derivative of the Digamma
function. 

\section{A toy model for the disorder}
\label{toymodel}

We employ here a toy model which facilitates the calculation of the
correlation function $C$ yet, we feel, captures some of the essential
physics.
This model has been previously employed in ref. \cite{sumrule} 
as part of our analysis of the transmission phase correlations through a
quantum dot. The model is based on the following simplifying assumptions:
\begin{itemize}
\item[(i)] The level spacing is constant, i.e. $\epsilon_j =\Delta
  \cdot j$. This mimics the level repulsion typical for a disordered
  system.
\item[(ii)] For the product of tunneling matrix elements we assume 
  $V_i ^L V_i^R=V^2 \alpha_i$ with $\alpha_i$ a 
  random variable which can
  take the values $+1$ and $-1$ with equal probability. This models
  the fluctuations in the wave function caused by the disorder.
\end{itemize}

With this model we calculate the correlation function of 
transmission amplitudes in the cotunneling regime
\begin{eqnarray}
C_t(N_1,x_1,N_2,x_2) &=& \langle t(N_1,x_1) t(N_2,x_2)^* \rangle
\end{eqnarray}
and show that we obtain the same results as with the formalism of
 section \ref{formal}.
For one disorder configuration (a sequence of the $\alpha_j$) the
transmission amplitude in the cotunneling regime (with $N$ electrons) 
at low temperature $kT\ll\Delta$ becomes
\begin{eqnarray}
t &=& V^2 \left ( \sum_{j\le N}
  \frac{\alpha_j}{E-((k-M)\Delta -xU)+i\Gamma} +\right. \nonumber\\
& & \left. + \sum_{j\ge M+1}
  \frac{\alpha_j}{E-((k-M)\Delta +U(1-x)) +i\Gamma} \right). 
\label{toytrans}
\end{eqnarray}
The disorder average is now equivalent to an
average over the $\alpha_i$. We find $\langle \alpha_i \rangle=0$ 
and $\langle \alpha_i \alpha_j \rangle =\delta_{i,j}$. 
We set $E=0$ and obtain 
\begin{eqnarray}
& & C_t(N,x_1,N+M,x_2) = \frac{V^2}{\Delta^2}
\left ( \sum_{k=1}^{N}\frac{1}{k-N-x_1 {\cal U}}\frac{1}{k-N-M-x_2
    {\cal U}} \right . \nonumber  \\
 & & +\sum_{k=N+1}^{N+M} \frac{1}{k-N+{\cal U}(1-x_1)}
 \frac{1}{k-N-M-x_2 {\cal U}} \nonumber\\
& & \left. +\sum_{k=N+M+1}^{\infty} \frac{1}{k-N+ {\cal U}(1-x_1)}
\frac{1}{k-N-M+{\cal U}(1-x_2)} \right),
\end{eqnarray}
where ${\cal U}=U/\Delta$. The sums can be expressed in terms of
the Digamma function $\psi$
\begin{eqnarray}
\sum_{n=0}^{M-1}
  \frac{1}{(n+a)(n+b)}& =& \frac{1}{b-a}(\psi(b)
  -\psi(a)+\psi(M+a)-\psi(M+b)) \ .
\end{eqnarray}
We arrive at
\begin{eqnarray}
& &\frac{\Delta^2}{V^2} C_t(N,x_1,N+M,x_2) = 
\frac{1}{M+(x_2-x_1){\cal U}} \left [ \psi(M+{\cal U}x_2)-\psi({\cal
  U} x_1) \right . \nonumber\\
& & \left . + \psi(N+{\cal U} x_1) - \psi(N+M+{\cal U}x_2) +
\psi(1+M+{\cal U} (1-x_1)) - \psi(1+{\cal U}(1-x_2)) \right]
\nonumber\\
& &+\frac{1}{M+(1+x_2-x_1){\cal U}} \left [ \psi(1+M+{\cal
  U}(1-x_1)-\psi(1+{\cal 
  U}(1- x_1) + \psi(M+{\cal U}x_2 - \psi({\cal U}x_2) \right]. \ \ \ \
\ \ \ \ \ \
\end{eqnarray}
When we are sufficiently far away from the resonances $x_{1,2} {\cal
  U}\gg1$ we can use the asymptotic expansion of the Digamma 
function $\psi(z)\sim \log z$ for $z\gg 1$. In the limit $N\rightarrow
  \infty$ we obtain
\begin{eqnarray}
& &  \frac{\Delta^2}{V^2} C_t(N,x_1,N+M,x_2)=
\frac{1}{M+(x_2-x_1){\cal U}} 
\log\left [\frac{M+x_2 {\cal U}}{x_1 {\cal U}} \frac{1+M+(1-x_1){\cal
    U}}{1+(1-x_2){\cal U}} \right] +\nonumber\\
& &+ \frac{1}{M+(1+x_2-x_1){\cal U}} \log \left[ \frac{M+x_2{\cal U}}{x_2{\cal
    U}}\frac{1+M+(1-x_1){\cal U}}{1+(1-x_1){\cal U}}\right ] \, .
\label{toyresult}
\end{eqnarray}
With the formalism of section \ref{formal} the correlation function
$C_t$ is given by 
\begin{eqnarray}
C_t(N_1,x_1,N_2,x_2)&=& \Delta \frac{\hbar}{e^2}{\cal G}^L
\frac{\hbar}{e^2} {\cal G}^R A(N_1,x_1,N_2,x_2)
\end{eqnarray}
with $A$ defined in (\ref{defa}) and calculated in (\ref{ttstar}).
Comparison with (\ref{toyresult}) shows that \\
$V^2=\Delta^2 (\hbar {\cal G}^L)/e^2
(\hbar {\cal G}^R)/e^2$.
\end{appendix}

\end{document}